\documentclass[10pt,twocolumn]{ieee}
\usepackage{paralist}
\usepackage{mathptmx}
\usepackage{graphicx}
\usepackage{setspace}
\usepackage{latex8} 
\usepackage{fancyhdr}
\newcommand{\CommunitySuperNode}{LeafSuperNode}
\newcommand{\SuperNode}{SuperNode}

\setdefaultleftmargin{0.0em}{}{}{}{}{}
\title{SuperGraph Visualization}

\author{Jose Rodrigues, Agma Traina, *Christos Faloutsos and Caetano Traina\\ \ Depto. de Ci\^encias da Computa\c{c}\~ao\ \ \ \ \ \ \ \ \ *Computer Science Department\ \\ Inst. de Ci\^encias Matem\'aticas e de Comput.\ \ \ School of Computer Science\ \ \ \ \ \ \\ \ \ \ \ Universidade de S\~ao Paulo\ \ \ \ \ \ \ \ \ \ \ \ \ \ \ \ \ \ \ \ \ \ Carnegie Mellon University\ \ \ \ \\ CP 668-13560-970 S\~ao Carlos, SP, Brazil\ \ \ \ \ 5000 Forbes Ave - 15213-3891, USA\\ \{junio, agma, caetano\}@icmc.usp.br\ \ \ \ \  \ \ \ \ \ \ \ \ \ \ \ \ \ christos@cs.cmu.edu\ \ \ \ \ \ \\\
{\bf IEEE Copyright - http://ieeexplore.ieee.org/xpls/abs\_all.jsp?arnumber=4061172}
\\
}
\begin{document}

\maketitle
\thispagestyle{empty}

\begin{abstract}
\noindent{Given a large social or computer network, how can we visualize it, find patterns, outliers, communities? Although several graph visualization tools exist, they cannot handle large graphs with hundred thousand nodes and possibly million edges. Such graphs bring two challenges: interactive visualization demands prohibitive processing power and, even if we could interactively update the visualization, the user would be overwhelmed by the excessive number of graphical items. To cope with this problem, we propose a formal innovation on the use of graph hierarchies that leads to GMine system. GMine promotes scalability using a hierarchy of graph partitions, promotes concomitant presentation for the graph hierarchy and for the original graph, and extends analytical possibilities with the integration of the graph partitions in an interactive environment.
}
\end{abstract}

\vspace{-0.3cm}
\section{Introduction}
\vspace{-0.3cm}
\label{sec:Introduction}



\noindent{Up-to-date applications have produced graphs on the order of hundred thousand nodes and possibly million edges (referenced from here on as large graphs). Large graphs are found in numerous real-life settings: web graphs (web pages pointing to others with hypertext links), computer communication graphs (IP addresses sending packets to other IP addresses), recommendation systems, who-trusts-whom networks, bipartite graphs of web-logs (who visits which page), to name a few. For such domains, efficient graph visualization becomes prohibitive due to the excessive processing power requirements that prevent interaction. Besides that, hundred-thousand-node drawings result in unintelligible cluttered images that do not aid to the user's cognition.}

To face these challenges, former works (section \ref{sec:RelatedWork}) propose to present large graphs based on a hierarchy of graph partitions. However, these efforts fail on the task of integrating the groups of nodes that constitute the levels of the hierarchy. In these propositions, the graph hierarchy is ``dead'' and cannot answer questions such as {\it What is the relation between a given group of nodes and another group of nodes? How many edges connect these two groups? Which are they? Which are the graph nodes from other groups that connect to a graph node of interest?} These questions translate to the possibility of using the original graph information concomitant to its hierarchical version. In such scenario, it is possible to benefit from both structurings in parallel or in cooperation. The main contribution of this work is the delineation of a system that can answer these questions dynamically and present them visually.

We review related works in section \ref{sec:RelatedWork}, in section \ref{sec:Terminology} we present basic concepts. Section \ref{sec:Connectivity} introduces new definitions for hierarchies of graph partitions and section \ref{sec:GraphTree} explains how to use these concepts in a suitable data structure. Section \ref{sec:Building} clarifies the construction of the data structure that supports our system and section \ref{sec:Experiments} presents the experiments. Section \ref{sec:Conclusions} concludes the paper.

\vspace{-0.3cm}
\section{Related Work}
\vspace{-0.3cm}
\label{sec:RelatedWork}

\noindent{In the literature, there are several works that deal with the problem of visualizing large graphs. Munzner \cite{45} proposes the H3 system, which deals with visual overload issues by using a specific spanning tree, and manages the scalability with an innovative dynamic hyperbolic layout. Different from this work, the system is based on a single resolution visual exploration and therefore has limited scalability features. Schaffer {\it et al} \cite{47} compare full-zoom navigation techniques and the fisheye view for drawing clustered graphs. Walshaw and Cross \cite{51} work on the issue of hierarchically partitioning a graph. While Eades and Feng \cite{50} (multilevel layout) and Frishman and Tal \cite{49} (dynamic layout), propose algorithms for determining the layout of clustered graphs. These works, though, have not considered the complete information of the groups of edges between the graph partitions and, instead of embedding the original graph in a supporting structure, the graph is lost if not kept in a parallel structure. Scalability is not considered in these works. These characteristics seriously limit their propositions, which lack interaction and data retrieval tasks.}

Papadopoulos {\it et al} \cite{42} propose to draw a graph based on the graph modular decomposition theory \cite{41}. Their work explores a recursive tree-like partition of a graph to draw different levels of a graph modular hierarchy. Their work is not a complete system, but a description of how to arrange the modules of a graph from different hierarchical levels. Interaction details are omitted. Eades \cite{43} also benefits from the recursive tree-like partition of a graph. His work proposes single resolution planar drawings that reflect the underlying structure of a clustered graph. His main motivation is improved aesthetics and not scalability.

Our system is based on any kind of hierarchy of graph partitions, which can be manually determined by the analyst or, for large graphs, can be automatically determined by a proper methodology. In our experiments, we apply a methodology named {\it k-way} partitioning. That is, given a graph $G = (V, E)$ with $|V|$ nodes and $|E|$ edges, we want to have $k$ subsets $V_1, V_2, . . ., V_k$ of $V$, such that $V_i \cap V_j = \emptyset$ for $i \neq j$, $|V_i| = n/k$ and $\cup _i V_i = V$. Also, the partitioning must minimize the number of edges of $E$ whose incident vertices belong to different subsets. This partitioning methodology is described by Karypis and Kumar\cite{1}.

\vspace{-0.3cm}
\section{Basic Terminology}
\vspace{-0.3cm}
\label{sec:Terminology}

\begin{figure}[htb]
	\centering	
\includegraphics[width=0.39\textwidth]{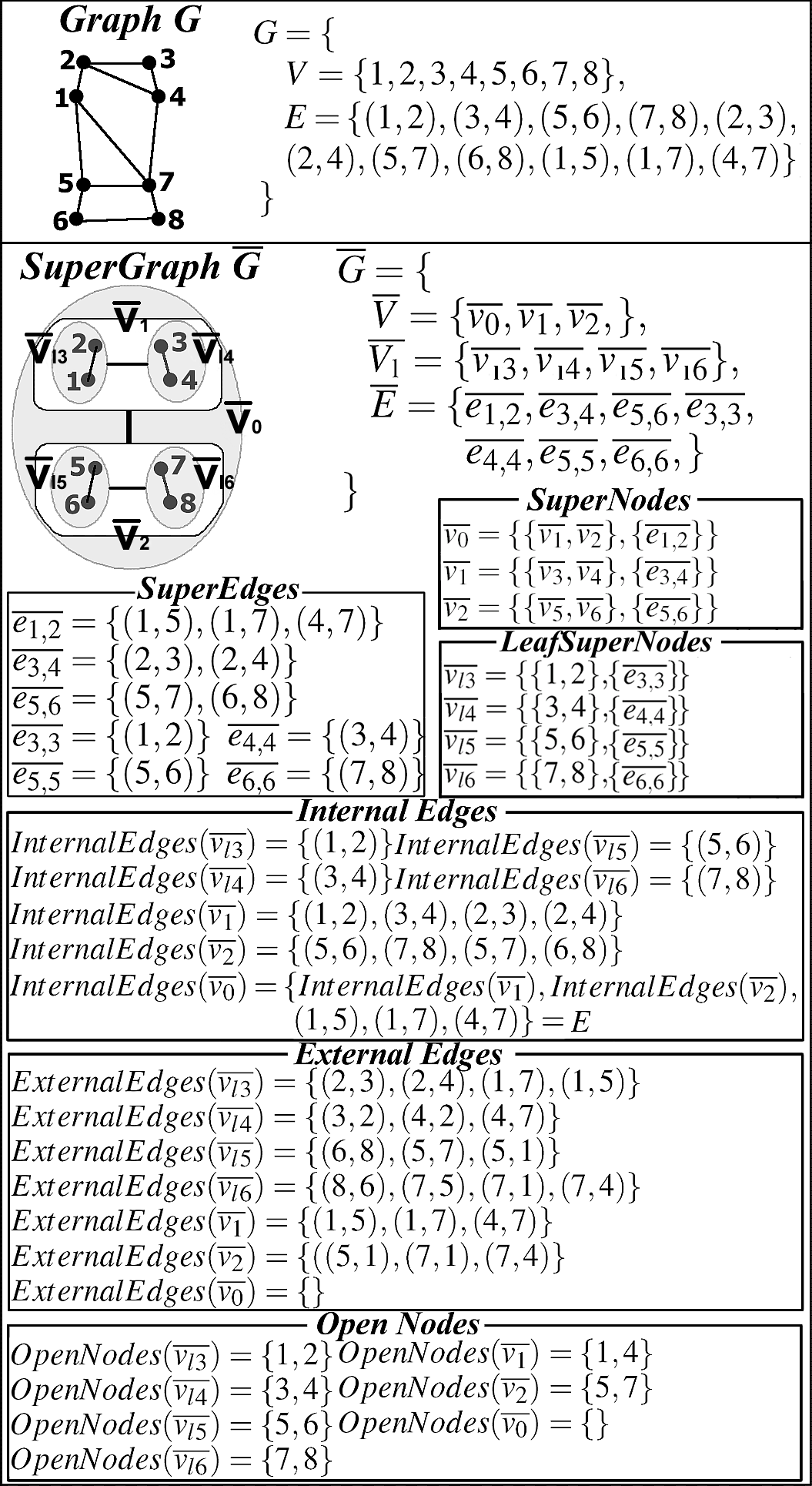}
	\caption{Example of a Graph and the respective SuperGraph.}
	\label{fig:GraphExample}
\end{figure}

\noindent{In this work, a hierarchy of graph partitions is called a SuperGraph. The underlying data beneath a SuperGraph is a Graph $G=\{V,E\}$, but a SuperGraph presents a different abstracting structure. It benefits from the fact that the entities and the relationships of the graph $G$ can be grouped according to the relationships that they define. In a SuperGraph, each of these groups of nodes is treated as a subgraph. This concept allows to work with a graph as a set of partitions hierarchically defined. Following, we define the constituents of a SuperGraph together with an illustrative example given in figure \ref{fig:GraphExample}}.\\

\noindent{\textbf{Graph and SuperGraph}}\\
Given a finite undirected graph $G=\{V,E\}$, with no loops nor multiple edges, a SuperGraph is a set $\overline{G}=\{\overline{V},\overline{V_l},\overline{E}\}$. More specifically, a SuperGraph is composed of a set $\overline{V}$ of {\SuperNode}s $\overline{v}$, a set $\overline{V_l}$ of {\CommunitySuperNode}s $\overline{v_l}$ and a set $\overline{E}$ of SuperEdges $\overline{e}$. Following we define {\CommunitySuperNode}, {\SuperNode} and SuperEdge.\\

\noindent{\textbf{{\CommunitySuperNode}s and {\SuperNode}s}}\\
Given a subset $V' \subset V$, a {\it \CommunitySuperNode} $\overline{v_l}$ is the subgraph $G|V'$ induced by $V'$. That is, $G|V'=\overline{v_l}=G'=\{V',E'\}$, $E' \subset E$. The set of {\CommunitySuperNode}s is totaly disjoint, that is:
\vspace{-13pt}

\begin{equation}
\bigcap{V'_i=\emptyset},
\ for \ \overline{v_{li}} \in \overline{V_l}\ and\ 
\ V'_i \in \overline{v_{li}}=\{V_i',E_i'\}
\label{eq:IntersectionOfSubgraphs}
\end{equation}

The union of the nodes of all the {\CommunitySuperNode}s of a SuperGraph equals to the set of nodes $V$. This fact is illustrated in the list of {\SuperNode}s in figure \ref{fig:GraphExample} and defined as follows:

\vspace{-13pt}

\begin{equation}
 \bigcup{V'_i=V},
\ for \ \overline{v_{li}} \in \overline{V_l}\ and\ 
\ V'_i \in \overline{v_{li}}=\{V_i',E_i'\}
\label{eq:UnionOfSubgraphs}
\end{equation}

A {\it {\SuperNode}} $\overline{v}$ is defined as a set of {\SuperNode}s $\overline{v}$ or, exclusively, it is defined as a set of {\CommunitySuperNode}s $\overline{v_l}$. Plus a set of SuperEdges $\overline{e}$. As follows:

\vspace{-13pt}

\begin{equation}
\begin{array}{c}
\overline{v}=\{\overline{V'}=\{\overline{v_0},\overline{v_1},...,\overline{v_{(|\overline{v'}|-1)}}\},\overline{E'}=\{\overline{e_{ij}}|
\{\overline{v_i},\overline{v_j}\} \subset \overline{V'}\}\}\cr
 OR \cr
\overline{v}=\{\overline{V_l'}=\{\overline{v_{l0}},\overline{v_{l1}},...,\overline{v_{l(|\overline{v_l'}|-1)}},\overline{E'}=\{\overline{e_{kk}}|\overline{v_{lk}} \in \overline{V_l'}\}\}\cr
\end{array}
\label{eq:SuperNode}
\end{equation}

Where the SuperEdge concept, $\overline{e}$, is defined further in this section.\\

\noindent{\textbf{Closure of a {\SuperNode}}}\\
In a SuperGraph, the closure of a {\SuperNode}, or {\CommunitySuperNode}, $\overline{v}$ is the set of all the graph nodes $v \in V$ that, ultimately, belong to {\SuperNode} $\overline{v}$. That is, given a {\SuperNode} $\overline{v}=\{\overline{V'},\overline{E'}\}$, the closure of $\overline{v}$ is given by the recursive definition:

\vspace{-13pt}

\begin{eqnarray}
Closure(\overline{v})=
\begin{cases}
V',\ if \ \overline{v}=\{V',E'\}\ is \ a \ \CommunitySuperNode \cr
\bigcup Closure(\overline{v_i}),\ for\ \overline{v_i} \ \in \ \overline{v},\ otherwise \cr
\end{cases}
\label{eq:Closure}
\end{eqnarray}

For example, in figure \ref{fig:GraphExample}, we have $Closure(\overline{v_2})=Closure(\overline{v_{l5}}) \cup Closure(\overline{v_{l6}})=\{5,6\} \cup \{7,8\}=\{5,6,7,8\}$. Also in the graph of figure \ref{fig:GraphExample}, $Closure(\overline{v_0})=V$. This last equality holds for any SuperGraph. The closure of a {\SuperNode} corresponds to the nodes that comprehend its community. Accordingly, at the lowest level of the tree (at the leaves) a community is a subgraph. At the highest level of the tree (at the root) the community is the entire graph. Intuitively, we refer to the parent of a SuperNode $\overline{w}$ as $Parent(\overline{w})=\overline{v}$ if $\overline{w} \in \overline{V'},\overline{V'} \in \overline{v'}=\{\overline{V'},\overline{E'}\}$. We refer to the set of parents of a SuperNode $\overline{w}$ as the set $Parents(\overline{w})=\{\overline{v}|\overline{w} \in Closure(\overline{v})\ and\ \overline{v} \in \overline{V}\}$.\\

\noindent{\textbf{SuperEdges}}\\
A SuperEdge $\overline{e_{ij}}$ corresponds to the SuperEdge for {\SuperNode}s $\overline{v_i}$ and $\overline{v_j}$. This SuperEdge holds the edges $e \in E$ that connect graph nodes from {\SuperNode} $\overline{v_i}$ to graph nodes from {\SuperNode} $\overline{v_j}$. A SuperEdge $\overline{e_{kk}}$ corresponds to the SuperEdge for {\CommunitySuperNode} $v_{lk}$. This SuperEdge holds the edges that interconnect graph nodes in {\CommunitySuperNode} $v_{lk}$ and corresponds to $E_k'$. That is, $\overline{e_{kk}} = E_k'$ for $\overline{v_{lk}}=\{V_k',E_k'\}$. Also, for a SuperEdge $\overline{e}$, $weight(\overline{e})=|\overline{e}|$; for an edge $e=\{v_p,v_q\}$, $source(e)=v_p$ and $target(e)=v_q$ (although we are assuming undirected graphs). Formally, a SuperEdge is defined as follows:

\vspace{-13pt}

\begin{equation}
\begin{array}{c}
   \overline{e_{ij}}=\{e| e \in E, \cr
   \ source(e) \in Closure(v_i)\ and\ target(e) \in Closure(v_j)\} \cr
\end{array}
\label{eq:SuperEdge}
\end{equation}

The union of the SuperEdges of all the SuperNodes together with the union of the SuperEdges of all the LeafSuperNodes equals to the set of edges $E$, as follows:

\vspace{-13pt}

\begin{equation}
\begin{array}{c}
                 ((\bigcup\overline{e_{ij}})
   \ \bigcup \
                  (\bigcup\overline{e_{kk}})) = E, \cr
\cr 
    \ for\ \{\overline{v_i},\overline{v_j}\} \subset \overline{V}\ and \ 
    \overline{v_{lk}} \in \overline{V_l}\cr
\end{array}
\label{eq:UnionOfSuperEdges}
\end{equation}

\vspace{-0.3cm}
\section{Extending the SuperGraph concept}
\vspace{-0.3cm}
\label{sec:Connectivity}

\noindent{The SuperGraph concept is a succinct representation for a hierarchy of graph partitions. However, hierarchies of graph partitions do not hold the original graph structure, which is inevitably lost when the hierarchical representation is used. In this section we define further concepts in order to extend the possibilities of a SuperGraph. Our aim is to answer the questions raised in section \ref{sec:Introduction} by dynamically restoring the original graph information.}\\

\noindent{{\bf Definition 1:}\ given a {\SuperNode}, or {\CommunitySuperNode}, $\overline{v}$, an edge $e$ is called an {\it internal edge} of $\overline{v}$ if $source(e) \in Closure(\overline{v})$ and $target(e) \in Closure(\overline{v})$. For this situation, we say that ``edge $e$ can be resolved within the closure of $\overline{v}$''. We define the set of all the {\it internal edges} of a {\SuperNode} $\overline{v}$ as $InternalEdges(\overline{v})$.}\\

\noindent{{\bf Definition 2:}\ an edge $e$ is called an {\it external edge} of $\overline{v}$ if $source(e) \in Closure(\overline{v})$ and $target(e) \not \in Closure(\overline{v})$. Accordingly, we say that ``edge $e$ cannot be resolved within the closure of $\overline{v}$''. We define the set of all the {\it external edges} of a {\SuperNode} $\overline{v}$ as $ExternalEdges(\overline{v})$.}\\

\noindent{{\bf Definition 3:}\ a graph node $v$, $v \in Closure(\overline{v})$, is an {\it open node} of $\overline{v}$ if there exists an {\it external edge} $e$, $e \in ExternalEdges(\overline{v})$, so that $source(e)=v$. We define the set of all the {\it open nodes} of a {\SuperNode} $\overline{v}$ as $OpenNodes(\overline{v})$.}\\

In the next subsections we explain how to use these definitions in order to extend the information that a SuperGraph can provide.

\subsection{SuperNodes Connectivity}
\label{subsec:sNodesConnectivity}
\noindent{We refer to the connections between groups of nodes in a graph hierarchy as {\it connectivity}. Formally, the connectivity corresponds to equation \ref{eq:SuperEdge}. According to the SuperGraph formalization, the connectivity for sibling communities is readily available as part of the SuperGraph, at its SuperEdges. For communities that are not siblings, or that are at different levels of the hierarchy, the connectivity must be traced.


The challenge here is how to trace the connectivity between arbitrary SuperNodes without having to cross the information of the hierarchy of partitions with the information of the underlying graph (not available). Instead, we are looking for a more scalable and efficient (viable) procedure for large graphs. In order to perform this task we use the {\it open nodes} information.

The {\it open nodes} information specifies all the nodes of a given {\SuperNode} that connect to nodes from other {\SuperNode}s.\\


\noindent{{\bf Theorem 1:}\ given any two {\SuperNode}s $\overline{v_i}$ and $\overline{v_j}$, the set of all possible edges connecting $\overline{v_i}$ to $\overline{v_j}$ is given by the Cartesian product $OpenNodes(\overline{v_i})\ X\  OpenNodes(\overline{v_j})$.\\

From equations \ref{eq:SuperNode} and \ref{eq:SuperEdge} follows that given a {\SuperNode} $\overline{v_f}=\{\overline{V'},\overline{E'}\}$, for any pair of sibling {\SuperNode}s $\{\overline{v_g},\overline{v_h}\},\ \{\overline{v_g},\overline{v_h}\} \subset \overline{V'}$, $\overline{v_f}$ is the unique {\SuperNode} that contains the edges connecting any pair of {\SuperNode}s $\{\overline{v_i},\overline{v_j}\},\ \{\overline{v_i},\overline{v_j}\} \subset \{Closure(\overline{v_g}) x Closure(\overline{v_h})\}$.\\

\noindent{{\bf Theorem 2:}\ the set of edges that actually connect any two {\SuperNode}s $\overline{v_i}$ and $\overline{v_j}$ is a subset of the unique SuperEdge $\overline{e_{gh}}$, which satisfies $\overline{v_i} \in Closure(\overline{v_g})$ and $\overline{v_j} \in Closure(\overline{v_h})|\  \{\overline{v_g},\overline{v_h}\} \subset \overline{v_f}$.\\

Intuitively, $\overline{v_f}$ is the first common parent of $\overline{v_i}$ and $\overline{v_j}$.

To determine the set of edges that connect any two {\SuperNode}s $\overline{v_i}$ and $\overline{v_j}$, we have to compute the intersection between the set of all possible nodes between $\overline{v_i}$ and $\overline{v_j}$ (theorem 1) and the set that contains the actual edges between $\overline{v_i}$ and $\overline{v_j}$ (theorem 2). That is:

\vspace{-13pt}

\begin{equation}
\label{eq:connectivity}
Connectivity(\overline {v_i } ,\overline {v_j } ) = \{ \begin{array}{*{20}c}
   \begin{array}{l}
 \{ OpenNodes(\overline {v_i } )\ X\ \\ 
 OpenNodes(\overline {v_j } )\}  \\ 
 \end{array}  \\
    \cap   \\
   \begin{array}{l}
 \{ \overline {e_{gh} } |\overline {v_i }  \in Closure(\overline {v_g } ), \\ 
 \overline {v_j }  \in Closure(\overline {v_h } )\}  \\ 
 \end{array}  \\
\end{array}\} 
\end{equation}

The SuperNode connectivity tells the relation between any pair of SuperNodes in a way that is possible to determine the number and which, exactly, are the graph nodes that determine the connectivity. This possibility extends the analysis for graph partitions because the SuperNodes are inspected either as sole entities or as groups of entities descending from the underlying graph.

\subsection{Graph Nodes Connectivity}
\label{subsec:gNodesConnectivity}
\noindent{A graph hierarchy uses the relationships among the graph nodes in order to define groups of related graph nodes. But the relationships between graph nodes at different groups of nodes are not part of the graph hierarchy. That is, the original graph is lost because the {\it external edges} information is not kept.}

In a SuperGraph, with the aid of the {\it open nodes} information, we can determine the complete set of {\it external edges} relative to any graph node.

From equation \ref{eq:SuperEdge} and definition 3, follows that given a graph node $v$, for any SuperNode $\overline{v}|v \in OpenNodes(\overline{v})$, there is one or more edges $e=\{v,w\}|e\in \overline{e}\ and\ \overline{e} \in Parents(\overline{v})$.\\

\noindent{{\bf Theorem 3:}\ if a graph node $v$ is an open node for a SuperNode $\overline{v}$, then the set of parents $Parents(\overline{v})$ have all the SuperEdges that hold edges connected to $v$.\\

Thus, if we know the set of parents and the set of open nodes of a SuperNode, we can determine the {\it external edges} of any graph node $v \in OpenNodes(\overline{v})$. To do so, a reference to the first parent SuperNode at each SuperNode is enough to define an incremental recursive procedure that can trace the external edges of any graph node of interest. Hence, while the graph node of interest is in the set of open nodes of the current parent SuperNode, there are still {\it external edges} to be traced. We just have to proceed upward in the hierarchy.

The graph node connectivity restores the original graph relationships dynamically. This way, in a million edges visualization, the user is guided across the hierarchy of partitions and allowed to inspect a particular node, instead of being overwhelmed by the huge volume of data.

\subsection{Integration to a data structure}
\noindent{The SuperGraph abstraction and the open nodes information define a novel structure model. This model provides a computational representation suitable to perform the operations defined in sections \ref{subsec:sNodesConnectivity} and \ref{subsec:gNodesConnectivity}.} In the next sections we illustrate the data structure used to implement this model. We explain how to build it at the same time that we gather the necessary information from the underlying graph.

\vspace{-0.3cm}
\section{Graph-Tree Structure}
\vspace{-0.3cm}
\label{sec:GraphTree}

\noindent{The Graph-Tree structure is intended to store and manage a SuperGraph. Since a SuperGraph is also a graph, the Graph-Tree is a new structure for graphs. Different from classic graph structures as adjacency matrices and lists of adjacencies, the Graph-Tree manages a graph according to a hierarchy of communities-within-communities. We explore this approach for large graph processing and visualization. To do so, the Graph-Tree is composed of {\SuperNode}s that are sets of {\SuperNode}s, and {\CommunitySuperNode}s that are sets of nodes. The later ones hold references to files storing subgraph information, one file per {\CommunitySuperNode}. These subgraphs are loaded to (expand {\CommunitySuperNode} task) and released from (collapse {\CommunitySuperNode} task) memory just when necessary, allowing for compartmented processing and presentation.

\begin{figure}[htb]
	\centering	
\includegraphics[width=0.39\textwidth]{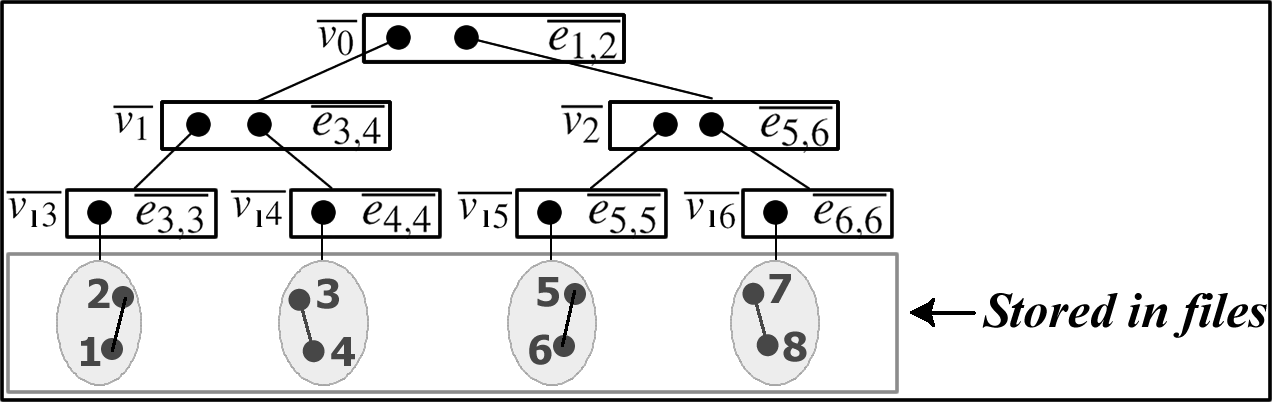}
	\caption{Graph-Tree example.}
	\label{fig:GraphTreeExample}
\end{figure}

For illustration, in figure \ref{fig:GraphTreeExample} we present the SuperGraph of figure \ref{fig:GraphExample} stored in a GraphTree. Notice how the tree adjusts to the {\SuperNode}s reflecting a hierarchical arrangement. For example, {\SuperNode} $\overline{v_2}=\{\overline{v_{l5}},\overline{v_{l6}}\}$ becomes parent of {\SuperNode}s $\overline{v_{l5}}$ and $\overline{v_{l6}}$. The SuperEdges are stored in the {\SuperNode}s' parent that holds references to the respective {\SuperNode}s. For example, {\SuperNode} $\overline{v_2}$ keeps the references of {\CommunitySuperNode}s $\overline{v_{l5}}$ and $\overline{v_{l6}}$, consequently, it holds SuperEdge $\overline{e_{5,6}}$. At the bottom of the tree we have subgraphs and their respective graph nodes and edges.\\

\noindent{\textbf{Components of the structure}}\\
To hold a SuperGraph, the Graph-Tree uses five sub structures to represent the concepts introduced in section \ref{sec:Terminology}: {\it open node} ({\it openNode}), edge ({\it edge}), SuperEdge ({\it sEdge}), LeafSuperNodes ({\it lNode}) and SuperNode ({\it sNode}). The first one, {\it openNode} is an alias for a node id, it refers to a node from a given community. The {\it edge} structure is used to abstract a relation (edge) between two nodes. The {\it sEdge} structure is used to abstract a set of edges between two {\SuperNode}s. The {\it lNode} structure represents a {\CommunitySuperNode} and the {\it sNode} structure represents a {\SuperNode}. Figure \ref{fig:GraphTreeComponents} details and exemplifies each of these structures.

\begin{figure}[htb]
	\centering	
\includegraphics[width=0.44\textwidth]{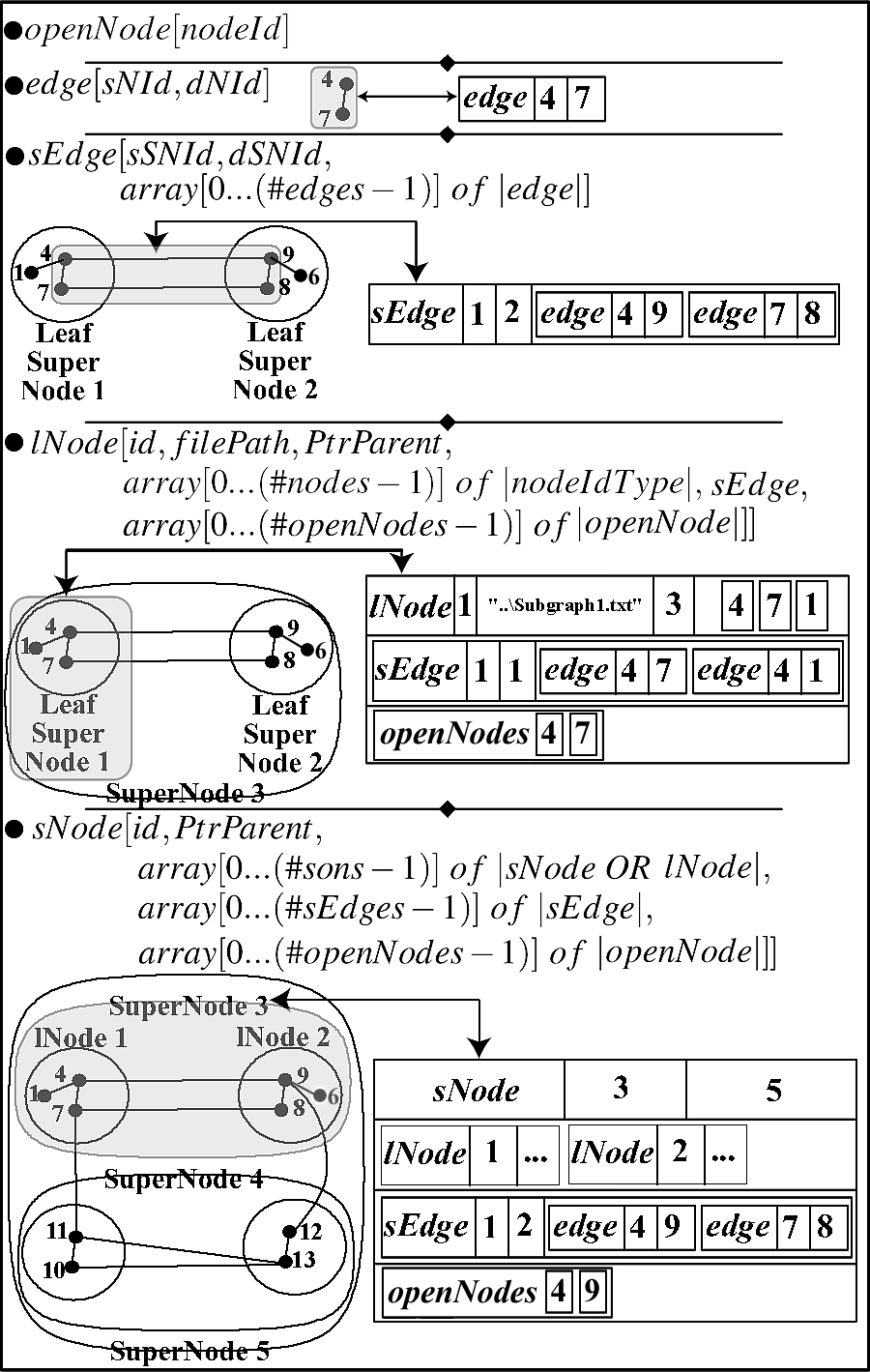}
	\caption{Graph-Tree components.}
	\label{fig:GraphTreeComponents}
\end{figure}

\vspace{-0.3cm}
\section{\textbf{Building a Graph-Tree from a graph}}
\vspace{-0.3cm}
\label{sec:Building}

\noindent{In this section we describe how to build a Graph-Tree departing from a graph. We illustrate all the steps in order to explain the process and to clarify the Graph-Tree structure, its arrangement and managed information.}\\


\noindent{\textbf{Hierarchy construction}}\\
Given a graph G=\{V,E\}, we recursivelly apply the k-way partitioning (section \ref{sec:RelatedWork}). We perform a sequence of recursive partitionings to achieve a hierarchy of communities-within-communities. At each recursion, each partition is submitted to a {\it k-way} partitioning cycle that will create another set of $k$ partitions. These partitions are propagated to the next level of the tree and the process repeats until we get the desired number of $h$ hierarchy levels. For each new set of partitions, a new subtree is embedded in the Graph-Tree structure and the references for the graph nodes are kept at the bottom level of the tree.\\

\noindent{\textbf{Filling the Graph-Tree {\SuperNode}s}}\\
After building the tree hierarchy based on the recursive partitioning, it is necessary to fill the {\SuperNode}s of the tree with the SuperEdge and {\it open nodes} information. In algorithm \ref{alg:TreeFilling}, we benefit from the tree structure to recursively scan the levels of the tree in a bottom-up fashion. Initially the LeafSuperNodes are filled with information from the subgraphs produced by the partitioning procedure. Then, we proceed to upper levels where the {\SuperNode}s use the {\it external edges} information propagated from lower levels.

\begin{algorithm}[htb]
  \caption{\emph{Algorithm to fill a Graph-Tree.}}
  \label{alg:TreeFilling}
  \begin{algorithmic}[1]
  \footnotesize
  \raggedright
  \REQUIRE $Ptr$: pointer to the root of the Graph-Tree\\
  \STATE $FillGraphTree(Ptr)$
  \STATE \textbf{if} $Ptr$ is leaf \textbf{then}
  \STATE \ \ Load subgraph file pointed by $Ptr$->$	filePath$.
  \STATE \ \ Instantiate and fill the $SuperEdge$ array of $edges$ and the array of son nodes for $Ptr$.
  \STATE \textbf{end if}
  \STATE \textbf{else}
  \STATE \ \ \textbf{for} each son $s_i$ of $Ptr$ \textbf{do}
  \STATE \ \ \ \ $FillGraphTree(s_i)$ /*Recursively down the hierarchy*/
  \STATE \ \ \textbf{end for}
  \STATE \ \ Instantiate a SuperEdge for each pair of sons.
  \STATE \ \ Use the external edges information to look for cross references between sons.
  \STATE \ \ Store resolved edges in the SuperEdges.
  \STATE \textbf{end else}
  \STATE Use external edges to determine $Ptr$'s array of open nodes.
  \STATE Propagate external edges information to parent.  
  \end{algorithmic}
\end{algorithm}

Figure \ref{fig:AlgorithmIllustration} illustrates this process. We start with graph $G$, which is partitioned to create the graph-tree with empty SuperNodes (see figures \ref{fig:AlgorithmIllustration}(a), \ref{fig:AlgorithmIllustration}(b) and \ref{fig:AlgorithmIllustration}(c)). The bottom-up recursive process starts at the leaves, illustrated in figure \ref{fig:AlgorithmIllustration}(d). For this illustration, and for figure \ref{fig:AlgorithmIllustration}(e), matches between {\it external edges} are indicated in boldface and gray {\it external edges} indicate unresolved {\it external edges}. Underlined nodes ids indicate {\it open nodes} and the diagonal arrows depict the {\it external edges} propagated up the tree. Still in figure \ref{fig:AlgorithmIllustration}(d), it is possible to see the information propagated from nodes $\overline{v_{l3}}$ and $\overline{v_{l4}}$, which will be used in step 11 of algorithm \ref{alg:TreeFilling} to find matches between unresolved external edges. Illustrated in figure \ref{fig:AlgorithmIllustration}(e), the crossing of the propagated information results in matches $(2,3)-(3,2)$ and $(2,4)-(4,2)$, stored in SuperEdge $\overline{e_{3,4}}$. Figure \ref{fig:AlgorithmIllustration}(e) also shows the first SuperEdges among siblings ($\overline{e_{3,4}}$ and $\overline{e_{5,6}}$) and another information propagation way up the tree. Figure \ref{fig:AlgorithmIllustration}(f) shows the last SuperEdge storing the last set of edges between siblings. Figure \ref{fig:AlgorithmIllustration}(g) shows the end of the process when no information is left for processing.\\

\begin{figure}[htb]
	\centering	   
\includegraphics[width=0.41\textwidth]{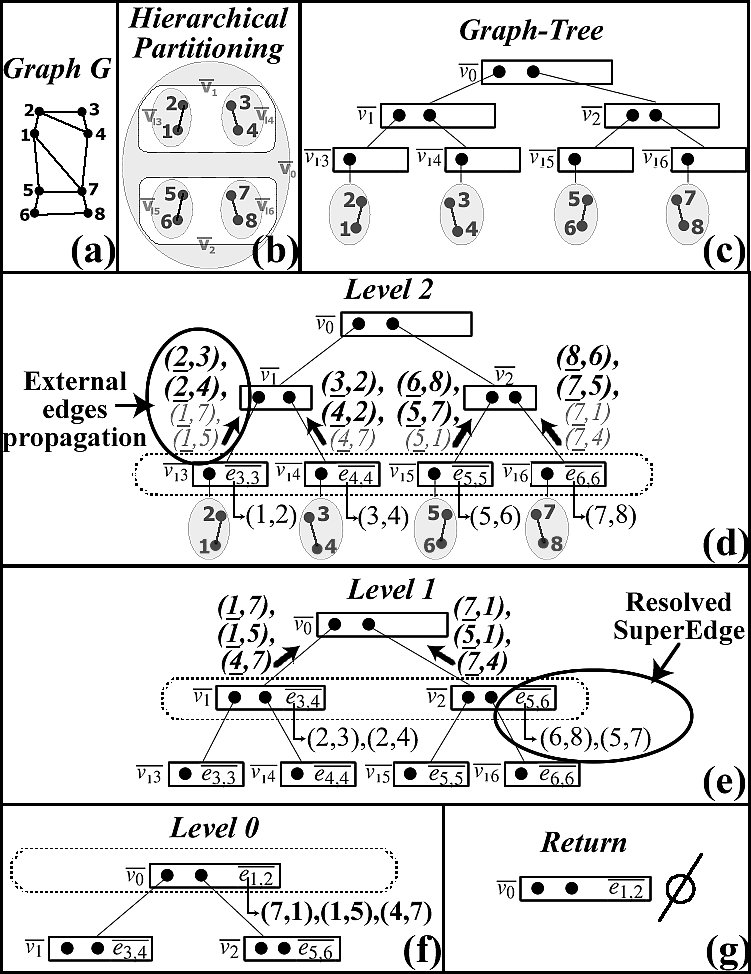}
	\caption{Graph-Tree filling illustration. From (a) to (c), hierarchical partitioning and empty Graph-Tree creation. From (d) to (g), illustration of the algorithm used to fill the Graph-Tree.}
	\label{fig:AlgorithmIllustration}
\end{figure}

\vspace{-0.3cm}
\section{Experiments with GMine}
\vspace{-0.3cm}
\label{sec:Experiments}

\vspace{-6pt}

\noindent{GMine implements the partitioning of a graph and manages this partitioning via integrated compartments. To do so, we use the Graph-Tree structure offering a set of interactivity tasks to visually mine a SuperGraph. Following, we illustrate the functionalities of GMine utilizing two datasets. Due to space limitations it is not possible to show all the GMine functionalities. Therefore, we have GMine available online at {\it \verb+http://www.cs.cmu.edu/~junio/GMine+}, where the software, datasets and videos can be downloaded.}\\

\noindent{\textbf{Email-net dataset}}

\noindent{The first dataset, which is intentionally small, defines a semantic-rich partitioning that was manually set in order to introduce the cognitive characteristics of GMine. It is comprised of $81$ nodes and $341$ edges. Each node represents an employee that belongs to a distinct company department. In the first level, the employees are grouped according to their department and in the second level according to their company, see figure \ref{fig:SyntheticDataset}(a). Each undirected edge of the graph represents electronic messages transmitted between two nodes, the weights indicate the number of messages. The visual interpretation of this graph aims at presenting the interrelationship between the individuals, the departments and/or the companies. This interrelationship is depicted by the number of messages exchanged between the entities of the SuperGraph.}

\begin{figure}[htb]
	\centering	
\includegraphics[width=0.47\textwidth]{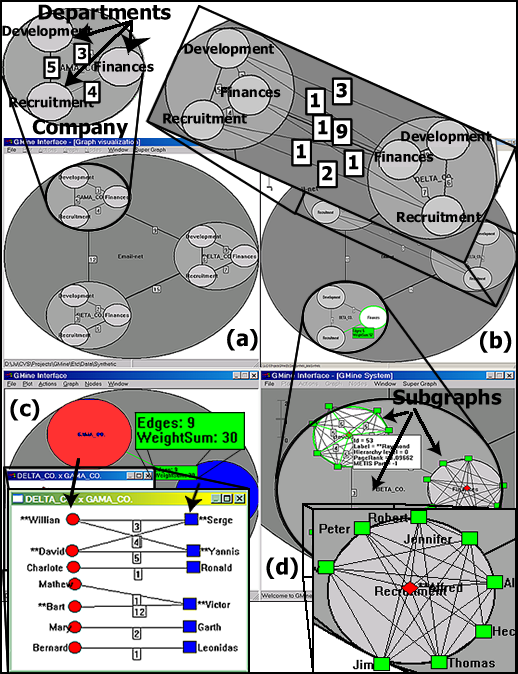}
	\caption{GMine visual mining of a synthetic dataset.}
	\label{fig:SyntheticDataset}
\end{figure}

We first illustrate relationships between {\SuperNode}s in figure \ref{fig:SyntheticDataset}(a). In this illustration we present SuperEdges among companies and SuperEdges among departments of the same company. Using equation \ref{eq:connectivity} and under user demand, we can calculate the relationship between departments of different companies, highlighted in figure \ref{fig:SyntheticDataset}(b). On top of the Graph-Tree structure, GMine system can also track the specific nodes (individuals) that exchange messages between two communities. On a double click event, the system presents these nodes (color differentiated) as a detailed bipartite subgraph in a separate window, highlighted in figure \ref{fig:SyntheticDataset}(c). In GMine, each subgraph can be processed totally independent of the rest of the visualization, including a set of graph processing tasks (sampling, partitioning, force-directed or page-rank based layouts), graph metrical calculations (degree distribution, components summary, hops) and rich interaction. Figure \ref{fig:SyntheticDataset}(d) shows how GMine permits to dig down the SuperGraph hierarchy and explore a specific community of nodes as a separate subgraph. It is possible to interact with a community subgraph in parallel to other community subgraphs, all in the context of the SuperGraph being visualized.\\

\noindent{\textbf{DBLP dataset}}\\
\noindent{The second dataset originates from the Digital Bibliography \& Library Project (or DBLP). DBLP is a publicly database of publication data that embraces authors from the Computer Science community and their published works, it is available at {\it \verb+http://dblp.uni-trier.de/+}.
The DBLP dataset version that we use defines a graph with $315,688$ nodes and $1,659,853$ edges, where each node represents an author from this community and each edge denotes a co-authoring relationship. In our experiments, we used GMine to automatically create a recursive partitioning of DBLP dataset. The partitioning has $5$ hierarchy levels each of them with $5$ partitions. The dataset, thus, is broken into $5^4 + 1$, or $626$, communities with an average of $500$ nodes per community. The communities reflect the connectivity among their members according to the k-way partitioning that, for this dataset, generates communities oriented to highly collaborative authors and consequentially to research themes.}

\begin{figure}[htb]
	\centering
	\label{fig:outlier}  
\includegraphics[width=0.47\textwidth]{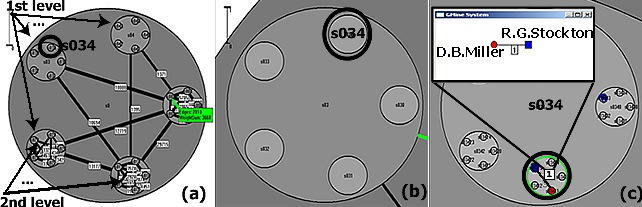}
	\caption{(a) Overview of DBLP dataset. (b) Focus on community {s34}. (c) Inspection of outlier.}	
\end{figure}

Figure 6 presents an overview of DBLP dataset. In figure 6(a), it is possible to see DBLP partitioned into $5$ communities in its first hierarchy level, and other $5*5$, or $25$ communities in its second hierarchy level. At this point, $3$ communities are highly connected to every other community and also highly connected among their $5$ sub communities. The other $2$ first level communities are relatively isolated from the other $3$ and totally isolated among their sub communities. One can conclude that the $3$ highly connected communities hold long term collaborating authors, while the other $2$ hold casual, less productive authors who seldom interact with each other. In figure 6(b) we focus on community {\it s034} and verify that its sub communities are isolated from each other. A deeper focus in community {\it s034} in figure 6(c) shows that among its sub communities (highlighted), only two of them present an edge. Our system allows to inspect this specific outlier edge to reveal that authors ``D. B. Miller'' and ``R. G. Stockton'' define this co-authoring relation for their unique DBLP publication dated from $1989$.

\begin{figure}[ht]
	\centering	   
\includegraphics[width=0.48\textwidth]{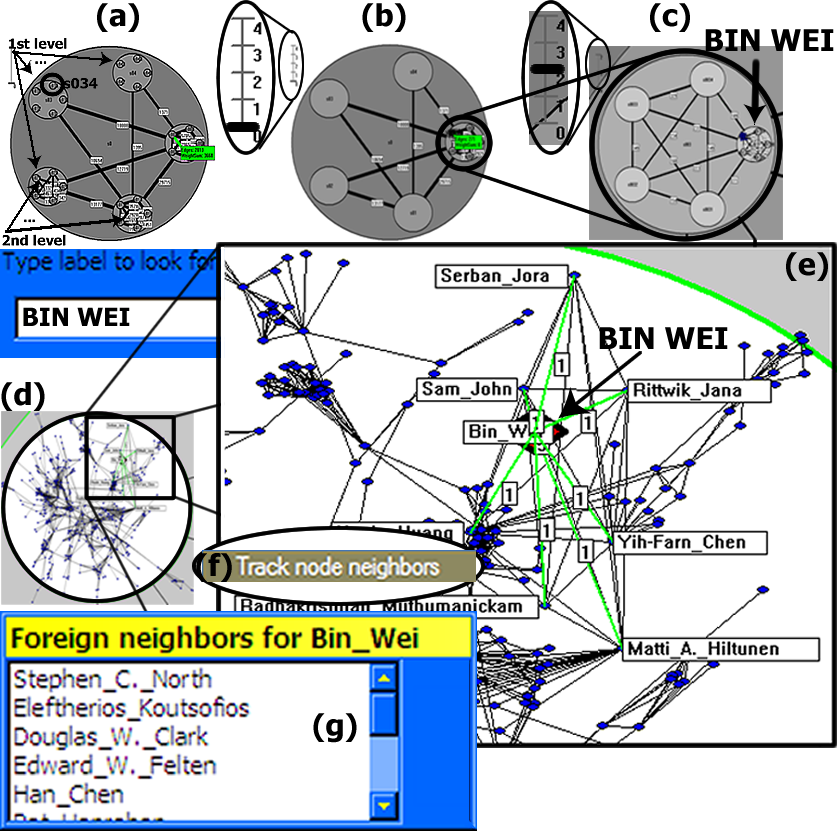}
	\caption{GMine visual mining of DBLP. (a) Query label for author Bin Wei. From (b) to (d), going down the graph hierarchy to find author Bin Wei. (e) Zoom on the subgraph of interest. (f) and (g), retrieval of external neighbors for the author of interest.}	
	\label{DBLPComplete}
\end{figure}

Figure \ref{DBLPComplete} presents a sequence of interactive actions performed by the user when navigating in DBLP dataset. Initialy, we perform a label query for author ``Bin Wei''. In figures \ref{DBLPComplete}(b) and \ref{DBLPComplete}(c), we illustrate the animation performed by GMine in order to show the graph node of interest. On the left-hand of the illustrations it is possible to see the level tracker indicating, at each step, the level of the hierarchy where the focus is. In figure \ref{DBLPComplete}(d) we reach the deepest level, where the subgraphs of the LeafSuperNodes are. Figure \ref{DBLPComplete}(e) zooms the direct relationships of author Bin Wei, which define a small community related to research on mobile computing authors. In figure \ref{DBLPComplete}(f) we use theorem 3 in order to retrieve the external neighbors for our sample author and get the list exhibited in figure \ref{DBLPComplete}(g). This list of authors indicate other communities where Bin Wei has research interest, including scientific visualization and distributed visualization.


\vspace{-0.3cm}
\section{Conclusions}
\vspace{-0.3cm}
\label{sec:Conclusions}
\noindent{We have presented GMine, a system for large graph visualization based on a hierarchy of graph partitions. We have covered the details to achieve our system by delineating and extending the SuperGraph concept and by introducing the Graph-Tree structure. We also demonstrated GMine using two datasets. In the experiments GMine was able to process and present different partitions of each dataset allowing targeted presentation under user's demand. The contribution of our work include scalability via partitioned processing and presentation of large graphs; extended analysis of a hierarchy of graph partitions by the integration of its parts in an interactive environment; and, most important, the possibility of concomitant functionalities for the hierarchy of graph partitions and the original graph.





\vspace{-0.3cm}
\section{Acknowledgements}
\vspace{-0.3cm}
\noindent{This work was partly supported by CAPES (Brazilian Committee for Graduate Studies), FAPESP (S\~ao Paulo State Research Foundation), CNPq (Brazilian National Research Foundation) and the National Science Foundation under Grants IIS-0209107, SENSOR-0329549 and IIS-0534205. This work was also partly supported by the Pennsylvania Infrastructure Technology Alliance (PITA) and by donations from Intel, NTT and Hewlett-Packard. Any opinions, findings, and conclusions or recommendations expressed in this material are those of the authors and do not necessarily reflect the views of the National Science Foundation, or other funding parties.
}


\vspace{-0.3cm}

\begin{spacing}{0.3}
\bibliographystyle{plain}

\end{spacing}

\end{document}